\documentclass[epj]{svjour}
\usepackage{latexsym}
\usepackage{graphics}
\usepackage{verbatim}

\usepackage{soul}

\usepackage{amsmath}
\usepackage{graphicx}
\usepackage{graphics}
\usepackage{physics}
\renewcommand{\figurename}{Fig.}
\usepackage[T1]{fontenc}
\usepackage{verbatim}
\usepackage[square,numbers,sort&compress]{natbib}
\usepackage[colorlinks=true, allcolors=blue]{hyperref}

\begin{document}
\title{Steering of Sub-GeV positrons by ultra-thin bent silicon crystal for ultra slow extraction applications}

\author{M. Garattini\inst{1,2} \and D. Annucci \inst{2,3} \and P. Gianotti\inst{1}  \and A. Liedl\inst{1}  \and E. Long\inst{1}  \and M. Mancini \inst{1,4} \and T. Napolitano \inst{1} \and M. Raggi \inst{2,3} \and P. Valente \inst{2}}


\institute{INFN Laboratori Nazionali di Frascati,Via E. Fermi 54, Frascati 00054, Italy \and INFN Roma 1, P.le Aldo Moro 5, Roma 00185, Italy \and  Physics Department, “Sapienza” Università di Roma, P.le Aldo Moro 5, Roma 00185, Italy \and Department of Physics, “Tor Vergata” Università di Roma,Via della Ricerca Scientifica 1, Roma 00133, Italy}

%
\abstract{
For the first time at the Beam Test Facility of the DA$\Phi$NE accelerator complex at the Laboratori Nazionali di Frascati of INFN, 450 MeV positrons have been deflected with high efficiency, using the Planar Channeling process in a bent silicon crystal. The deflection angle obtained is beyond 1 mrad. This interesting result finds several applications for manipulation of this kind of beams, in particular for slow extraction from leptons circular accelerators like DA$\Phi$NE. In this work the experimental apparatus, the measurement procedure and the experimental results are reported.}

%
%
\maketitle

\section{Introduction}
\label{sec:Intro}

Coherent processes of charged particles in bent single crystals are well known since 1970s, and in the last decades they found several application in the particle beam steering field.

A charged particle impinging on a bent crystal parallel with respect to a specific lattice plane orientation can oscillate in the potential well of the neighbours atom strings, being canalized following the crystal bending, and eventually being deflected. This process is commonly called ``Planar Channeling'' (CH). 

For a bent crystal, it is possible to define the critical CH angle \(\Theta_{c} \simeq \sqrt{(2U_{o}/pv)} \), where $U_{o}$ is the depth of the planar potential well; $p$ and $v$ are the particle momentum and velocity respectively. If the particle incident angle is smaller than $\Theta_{c}$ the particle can be captured and deflected by the CH effect \cite{biryukov2013crystal}.

Another coherent process, called ``Volume Reflection'' (VR), can happen in bent crystals. This reflection occurs when the particle trajectory is nearly tangential to one of the bent crystal planes. At the reflection point, the VR process occurs with large probability and the incoming particles are reflected to the opposite side with respect to the crystal bending \cite{scandale2007high}. 


Instead, when the particle incident angle is not in the range for CH or VR, the crystal behaves as a normal amorphous material and the particles are influenced by Multiple Scattering process. This condition is generally called ``Amorphous orientation'' (AM).

In the field of high energy hadrons, the CH process has been successfully studied and used for beam collimation, extraction and focusing applications \cite{scandale2016observation,afonin1998first,altuna1995high,Scandale2017SPSextraction,Fraser2017extraction1,Fraser:IPAC19-THXXPLM2}.

Instead, in the range of sub-GeV charged particles, and in particular for positrons, coherent processes in bent crystals are not completely explored and much less their application in accelerator physics. 

In this context, the SHERPA experiment at LNF-INFN investigated the possibility to use a bent silicon crystal to slowly extract 500 MeV positrons and electrons circulating in one of the DA$\Phi$NE accelerator complex \cite{garattini2021crystal}.

The last step to complete this feasibility study is the demonstration that it is possible to deflect positrons and electrons by an angle of the 1 mrad order using CH. In this work will be presented the experimental evidence of that, and in the case of positrons, the very first clear observation of CH in a bent crystal at sub-GeV energy.




\section{The Crystal}
\label{sec:TheCrystal}

The beam steering of sub-GeV positrons and electrons is very challenging from the point of view of the crystal and bending holder realization, especially if compared with multi-GeV hadrons applications. At these energies, the ``Dechanneling'' (DCH) process, which is competing with the CH one, is very effective due to the fact that silicon lattice electrons strongly interact with the impinging particles pushing them out of the CH condition \cite{biryukov2013crystal}. 

The DCH process follows an exponential behavior with a characteristic DCH length, of the order of some 100 $\mu$m at sub-GeV energies. This implies that it is necessary to use a crystal of some tens of $\mu$m along the beam direction to obtain a reasonable CH deflection efficiency. Previous experiments have been conducted in the past with 500 MeV positrons at the BTF using 1 mm thick silicon crystals, though they achieved low deflection efficiencies \cite{bellucci2006steering}

SHERPA silicon crystals are $\sim$ 15 $\mu$m thick and are mounted on a dynamic holder able to bend the crystal very precisely and uniformly by piezoelectric remote controlled motors. This technology has been developed by the INFN-Fe and LNL-INFN some years ago \cite{de2018innovative} and successfully used to deflect sub-GeV electrons \cite{mazzolari2014steering,wistisen2016channeling,sytov2017steering,bandiera2021investigation}. The SHERPA silicon crystal has been produced by Fe-INFN and the bending holder is an optimization based on the previous design. This new holder, realized by the SPCM (Servizio di Progettazione e Costruzioni Meccaniche) of LNF-INFN, results ten times cheaper and more engineered, especially about crystal mounting technologies and procedures, thus allowing a serial production of such bent crystals with an high repeatability and reliability.

\begin{figure}
\centering
\resizebox{0.3\textwidth}{!}{%
  \includegraphics{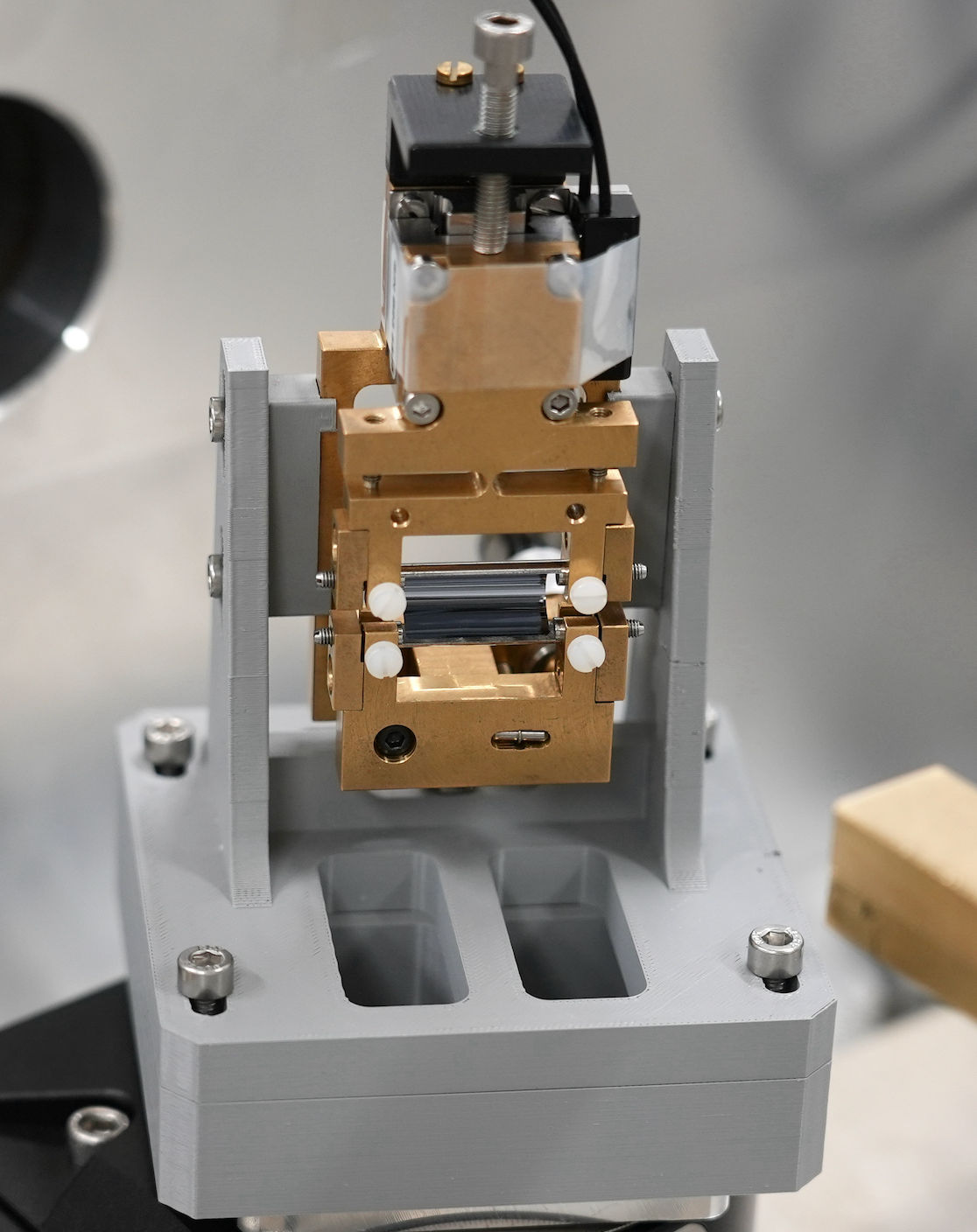}
}
\caption{The SHERPA silicon crystal bent by its dynamic holder.}
\label{Crystal_bent}       
\end{figure}

\begin{figure}
\centering
\resizebox{0.3\textwidth}{!}{%
  \includegraphics{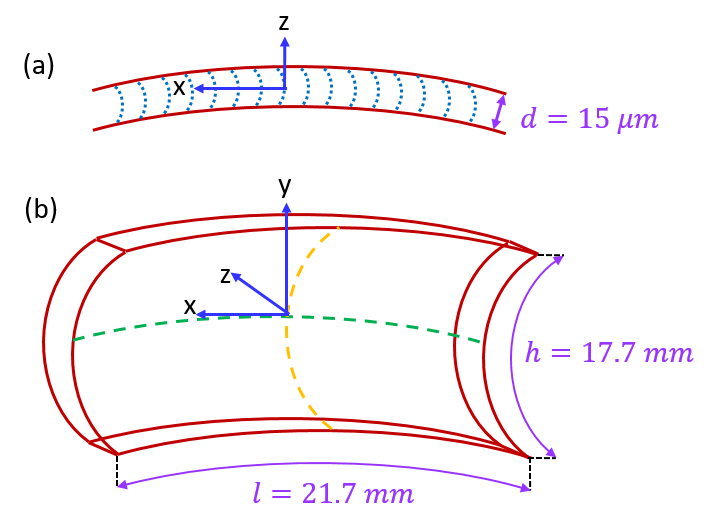}
}
\caption{SHERPA bent crystal scheme: in (a) the 2D cut-off of the crystal (111) planes (dotted blue lines). In (b) the 3D draw of the crystal bending, with respect to the beam axis, the z direction.}
\label{fig:ACQM}       
\end{figure}

In Fig.\ref{Crystal_bent} is shown a picture of the SHERPA crystal bent by its holder, while the crystal scheme and features are reported in Fig.\ref{fig:ACQM} .

The CH effect occurs along the (111) atomic planes, oriented parallel to the beam direction, bent by ``Quasi-mosaic'' (QM) curvature \cite{camattari2015thequasi} induced by the primary one visible in Fig.\ref{Crystal_bent} and Fig.\ref{fig:ACQM}.

An additional important advantage to use a 15 $\mu$m thick crystal is the possibility to obtain very small radius of primary curvature and, consequently, an high QM tertiary curvature, reaching more than 1 mrad of CH deflection. In fact, the silicon crystal results extremely elastic and deformable. At the same time, it is extremely fragile, making the bending techniques and procedures very challenging.

Before the beam characterisation at the ``Beam Test Facility'' (BTF) in Frascati, a bent crystal primary curvature of $R_P$ = (3.36 $\pm$ 0.01) mm has been measured by an optical profilometer. From this measurement it is possible to estimate quite precisely the QM tertiary curvature $R_{QM}$, along which CH occurs, and consequently the bending angle $\Theta_{b}$, by the following relations \cite{camattari2015thequasi}

\begin{equation}
R_{QM} = 3.539 \; R_{P}
\label{RQM}
\end{equation}
\begin{equation}
\Theta_{b} \sim t/R_{QM} 
\label{Tetha_def}
\end{equation}
where $t$ is the crystal thickness along the beam direction. The estimated bending angle is $\Theta_{b}$ = (1.26 $\pm$ 0.01) mrad.

\section{Experimental apparatus}
\label{sec:Apparatus}

The characterisation of the silicon bent crystal consists in the determination of its CH angle and efficiency. For this purpose it is necessary a particle beam spot with the smallest possible spot on the X-Y plane (Z is the beam propagation axis) and with the lowest divergence, especially on the deflection plane, the X-Z one in our case. Moreover, the alignment of the beam axis with respect to the geometrical center of the crystal is crucial.
The BTF beam emittance, coming directly from the DA$\Phi$NE LINAC, is not optimized for these specific measurements, so to reduce both dimension and divergence of the beam a collimator is needed. 

\begin{figure}
\centering
\resizebox{0.4\textwidth}{!}{%
  \includegraphics{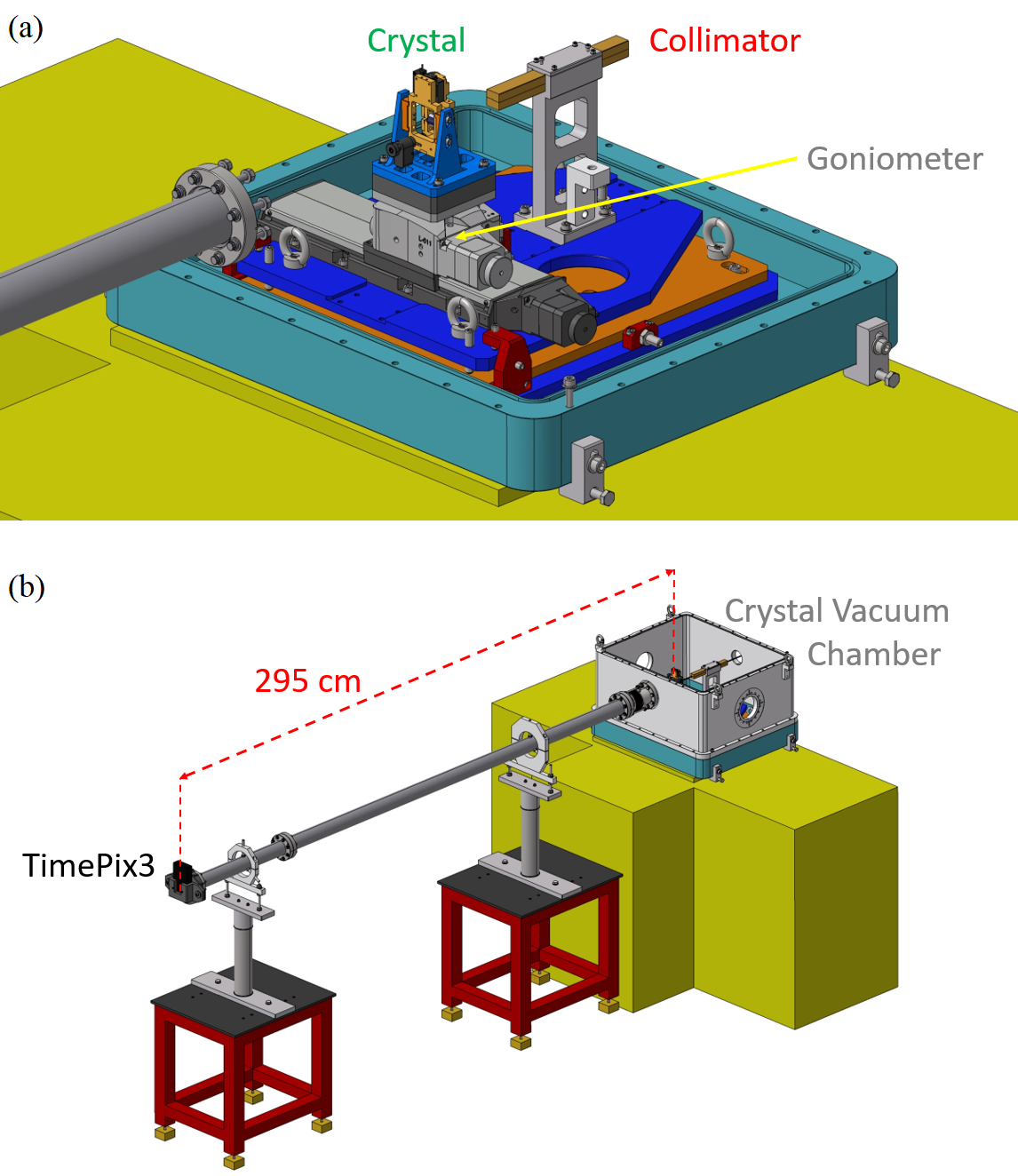}
}
\caption{SHERPA Experimental setup 3D CAD rendering: in (a) the goniometer, the collimator and the crystal position in the vacuum chamber are shown. In (b) the whole setup.}
\label{Sherpa_Apparatus_3d}       
\end{figure}

In Fig.\ref{Sherpa_Apparatus_3d} is shown the technical design of the whole apparatus. Upstream, the crystal chamber is mounted in front of the BTF end-line window ($\sim$ 150 $\mu$m of Titanium). In the chamber are mounted the 20 cm thick copper collimator, with a square aperture of (0.5 $\times$ 0.5) mm$^2$, and the crystal, mounted on a 3-axis movimentation system able to precisely orient the crystal with respect to the beam axis. All these elements, including the laser alignment tools, are very well mechanically pre-aligned all together at the 0.1 mm order. A 2D pixel detector (TimePix3 model by ADVACAM Company) is used to measure the particle distribution downstream the crystal.

To distinguish on the the TimePix3 detector the particle deflect by the crystal from all the others, it is necessary to mount it 3 m downstream the crystal and perform the measurements in vacuum to eliminate the multiple scattering of particles with the air \cite{annucci2023sherpa}. For the same purpose, extremely thin Mylar vacuum window ($\sim$ 50 $\mu$m) are used to close the vacuum chamber. The TimePix3 is mounted in air, just downstream the vacuum pipe window.

Using the BTF laser alignment system, all the SHERPA apparatus components have been pre-aligned with respect to the nominal beam axis. Then, by a beam based alignment procedure, it has been possible to optimize and maximize the flux coming out the collimator an passing through the crystal. An estimation of the beam transversal dimension at the crystal point are $\sim$ (0.5 $\times$ 0.5) mm$^2$ in X-Y plane and $\sim$ 2 mrad of divergence both in X and in Y.

\section{Measurements results}
\label{sec:Measuremets}

In December 2023, using the apparatus and the procedures described above, it has been possible to observe the CH and the VR processes of 450 MeV positrons and electrons in a bent silicon crystal. 

\begin{figure}
\centering
\resizebox{0.5\textwidth}{!}{%
  \includegraphics{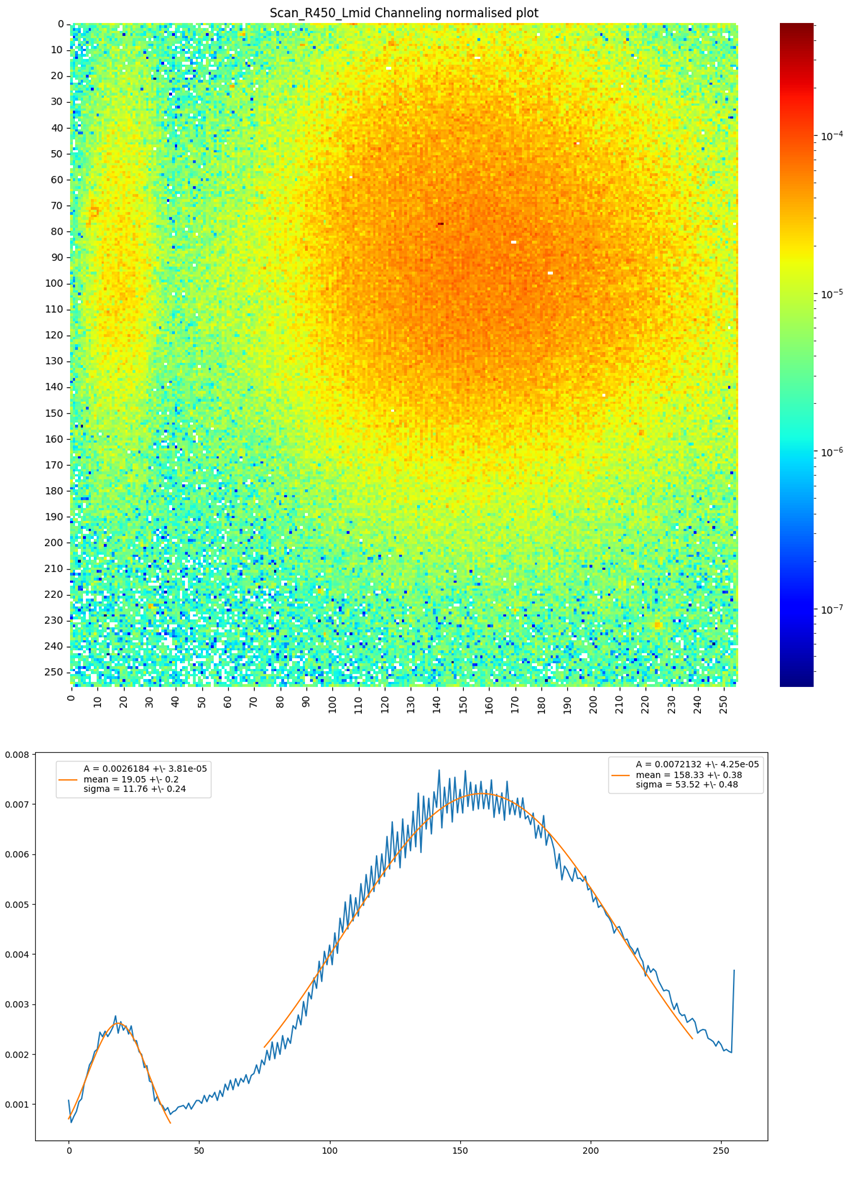}
}
\caption{Normalized X-Y Beam spot distribution of a positron Channeling run and corresponding X-projection profile.}
\label{fig:ChImg}       
\end{figure}

The particle distribution after the crystal, for different crystal planes orientation with respect to the beam axis, has been measured by the TimePix3 detector. Performing this angular scan it has been possible to find the best CH orientation, in which the deflected beam spot is more intense. In Fig.\ref{fig:ChImg} is shown the online TimePix3 2D beam image of this best CH orientation and its horizontal profile, integrating the intensity on the vertical one.

\begin{figure}
\centering
\resizebox{0.5\textwidth}{!}{%
  \includegraphics{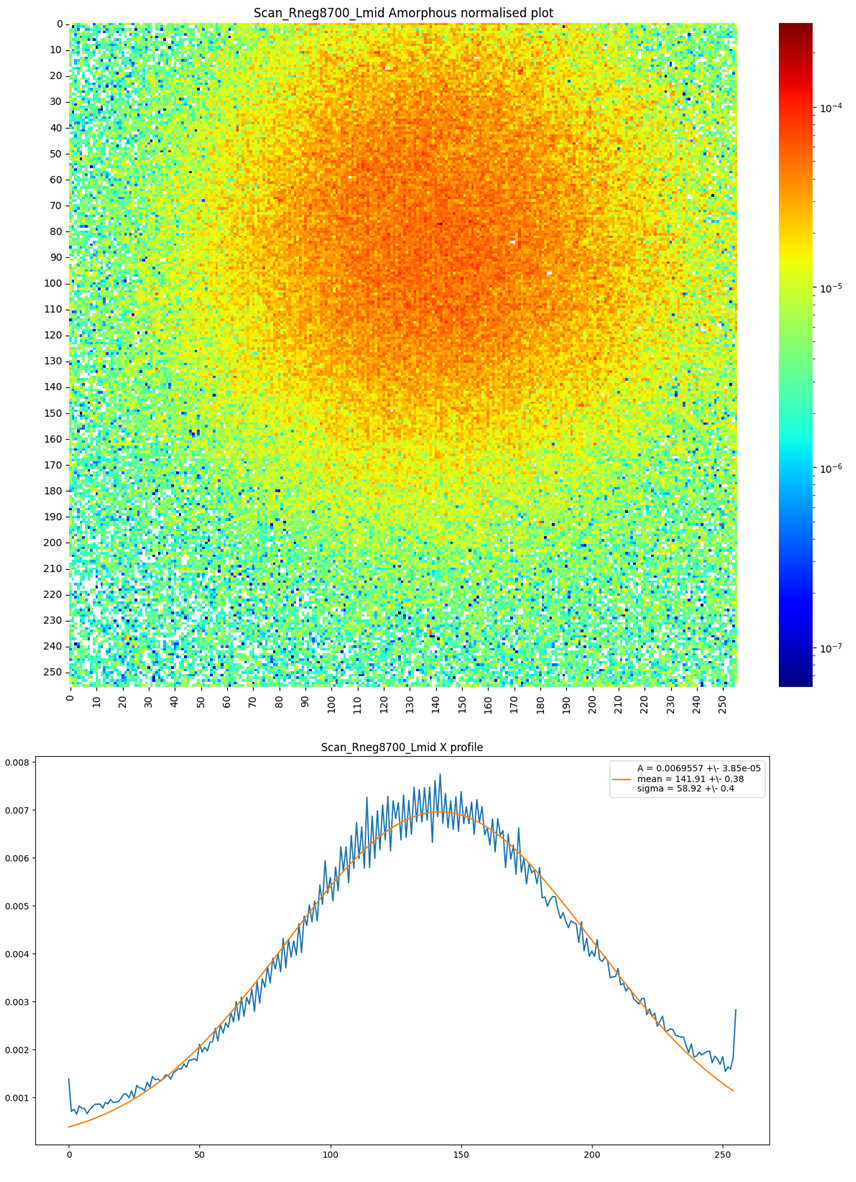}
}
\caption{Normalized X-Y Beam spot distribution of a positron Amorphous run and corresponding X-projection profile.}
\label{fig:AmImg}       
\end{figure}

\begin{figure}
\centering
\resizebox{0.5\textwidth}{!}{%
  \includegraphics{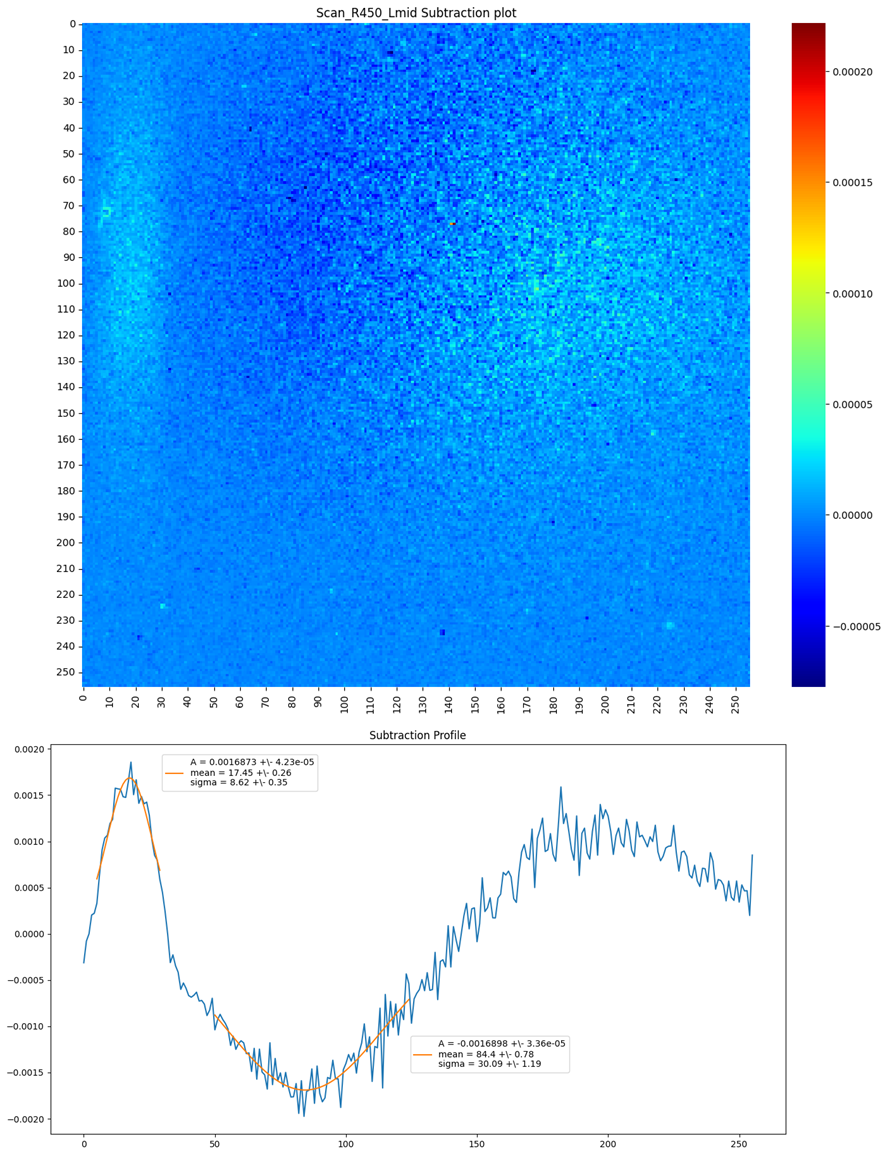}
}
\caption{Subtraction between the normalized positron CH and AM runs distributions and corresponding X-projection profile.}
\label{fig:CH-AmImg}       
\end{figure}

Since the beam divergence is much bigger than the channeling critical angle ($\Theta_c \sim300 \: \mu$rad), to distinguish the particles able to be channeled from all the others, it is necessary to apply a differential analysis subtracting to this distribution the distribution obtained with the crystal oriented in AM, applying the correct normalization. The AM distribution and its horizontal profile is shown in Fig.\ref{fig:AmImg}, while the difference between the CH and the AM distributions is reported in Fig.\ref{fig:CH-AmImg}.

\begin{figure}
\centering
\resizebox{0.4\textwidth}{!}{%
  \includegraphics{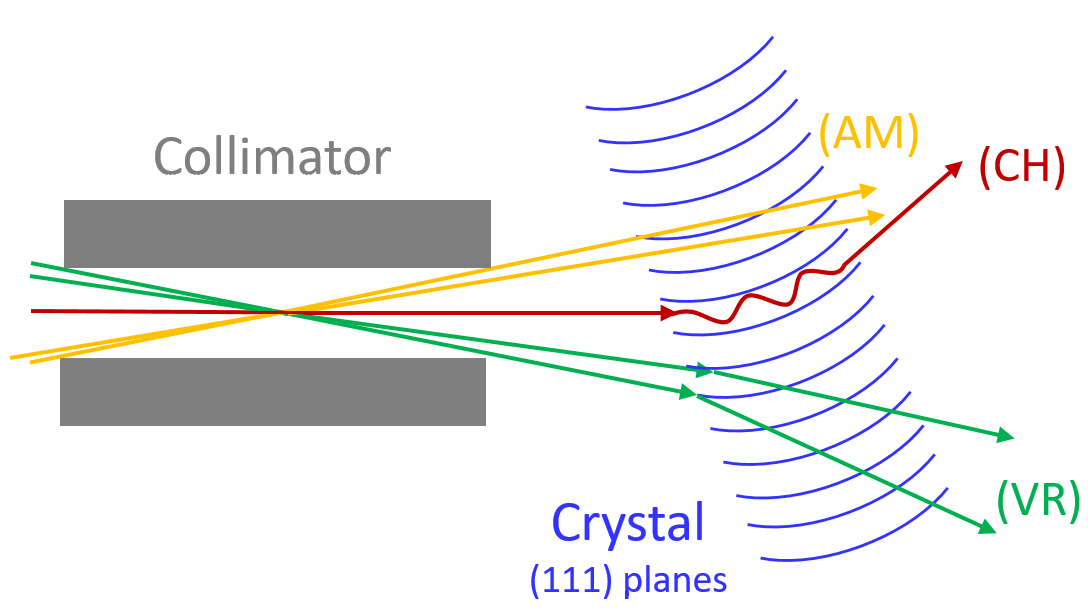}
}
\caption{Coherent processes in the SHERPA silicon bent Crystal.}
\label{SHERPACRY}       
\end{figure}

The large beam divergence causes different areas of the crystal, along the X axis, to have a different orientation relatively to the beam direction. This means that in the subtracted distribution appear the CH, the VR and AM effects at the same time. In Fig.\ref{SHERPACRY} it is shown a scheme of how the beam coming out from the collimator interacts with the crystal.

Doing a comparison between the distribution in CH and the AM one, it appears evident that some particles move from an area that we will call ``Emptying area'' to the CH and VR areas respectively.

The CH peak, on the left of Fig.\ref{fig:CH-AmImg}, is very evident as it is the well of the ``Emptying area'' in the center of the plot. In fact, the particles that hit the crystal in correspondence of the well X position are perfectly parallel to the silicon (111) planes and are deflected very efficiently by CH in the peak X position on the left side.

Consequently, the distance $d = (3.68 \pm 0.05)$ mm ($\sim$ 67 pixels) between the two peaks, positive and negative, will easily provide the bending angle of the crystal. In fact, considering that the crystal distance from the detector is $l=(2950 \pm 5)$ mm and that the pixel dimension is 55 $\mu$m, the bending angle is given by

\begin{equation}
\Theta_b = \arctan{(d/l)} = (1.23 \pm 0.02)\: \text{mrad}
\end{equation}

This result is extremely in good agreement with the $\Theta_{b}$ value of (1.26 $\pm$ 0.1) mrad estimated measuring the primary curvature of the crystal reported in  the previous section. 

It's important to underline that the crystal appears extremely effective deflecting positrons, and this is demonstrated by the fact that the amplitude ratio between the CH peak and the ``Emptying area'' well depth is $\sim$ 0.97. It means that, when the particles are very well aligned with the silicon planes, they are almost all deflected. This shows the very good quality of the crystal in terms of regularity and purity of the atomic lattice and homogeneity of its curvature.

Another clear evidence of the bending of the crystal is the presence of the VR positive plateau on the right of the Fig.\ref{fig:CH-AmImg} distribution, that shown a deflection of the beam on the opposite angle with respect to CH, as expected. The plateau is the result of the large angular acceptance of the VR process. Unfortunately, the present experimental conditions do not consent a quantitative and exhaustive study of VR process and additional dedicated measurements are needed.

\begin{figure}
\centering
\resizebox{0.5\textwidth}{!}{%
  \includegraphics{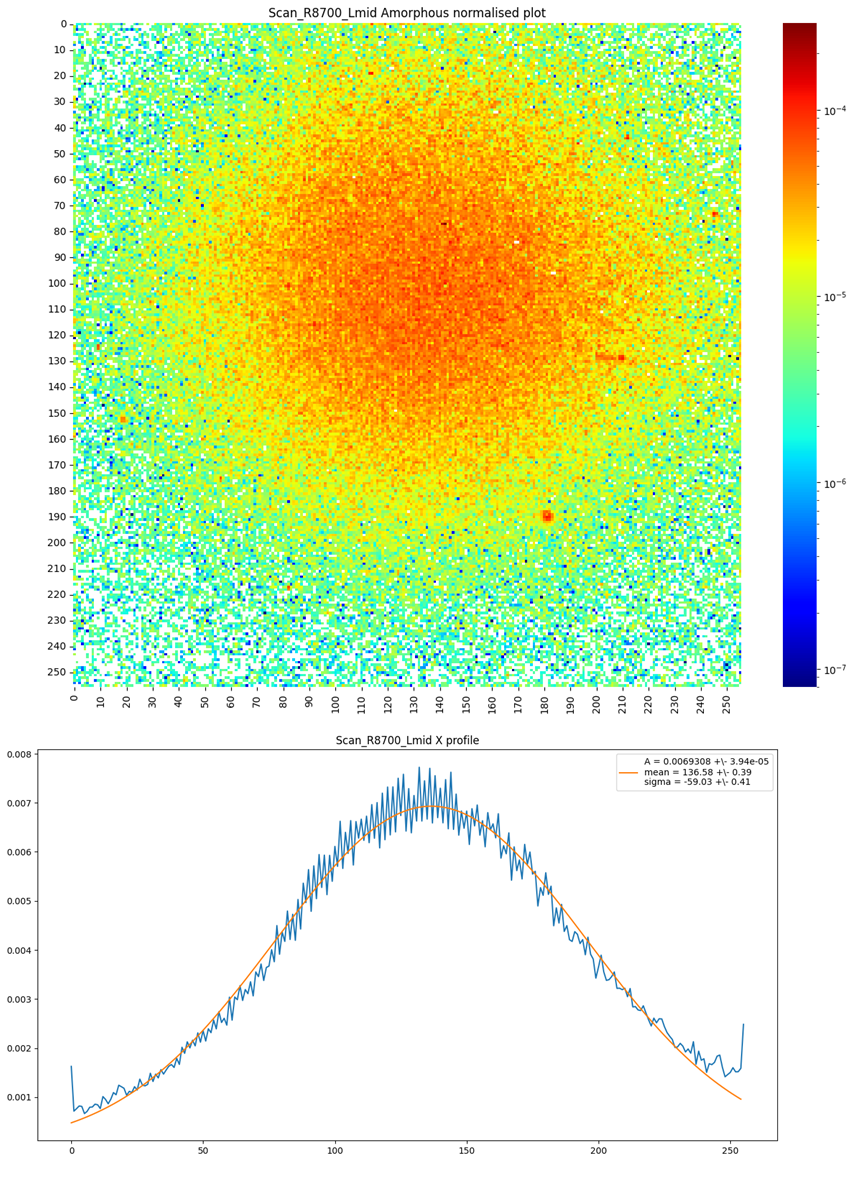}
}
\caption{Normalized 2D Beam spot distribution of the second positron Amorphous run and corresponding X-projection profile.}
\label{fig:AM2}       
\end{figure}

\begin{figure}
\centering
\resizebox{0.5\textwidth}{!}{%
  \includegraphics{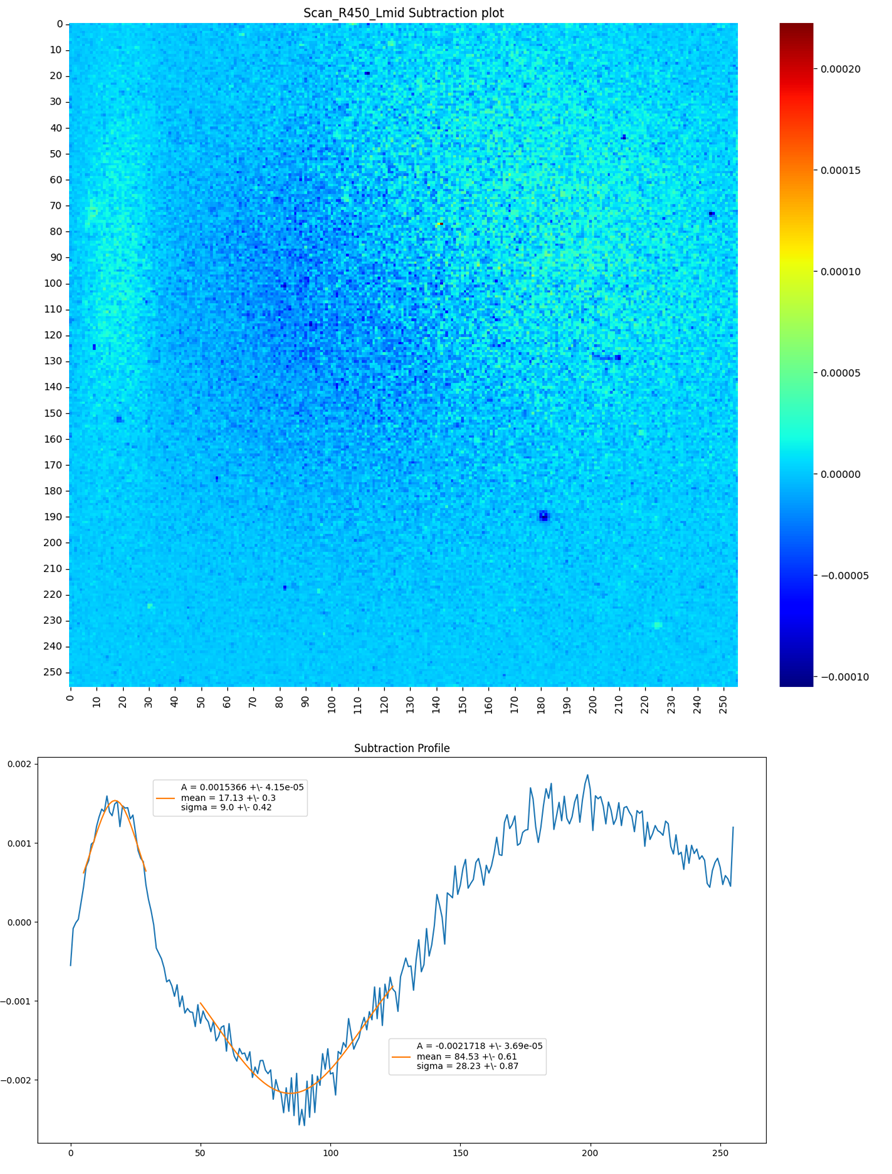}
}
\caption{Subtraction between the normalized positron CH and the new positron AM runs distributions and corresponding X-projection profile. }
\label{fig:CH2-AM2}       
\end{figure}

With the purpose to check the previous result, a different AM distribution, taken in a different moment and with a different number of total events (Fig.\ref{fig:AM2}), has been subtracted to the same CH distribution. The subtracted profile plot is reported in Fig.\ref{fig:CH2-AM2}. In this case the distance between the two peaks, positive and negative, amounts to \(d\) = (3.71 $\pm$ 0.05) mm, which corresponds to a beam deflection angle $\Theta_{b}$ = (1.26 $\pm$ 0.02) mrad. Again, the result is in extremely good agreement both with the previous measurement and the estimated value. The final error calculation on the $\Theta_{b}$ measured with the beam takes into account the uncertainty on the crystal-detector distance, the detector resolution, and the Gaussian fit errors. 



\begin{figure}
\centering
\resizebox{0.5\textwidth}{!}{%
  \includegraphics{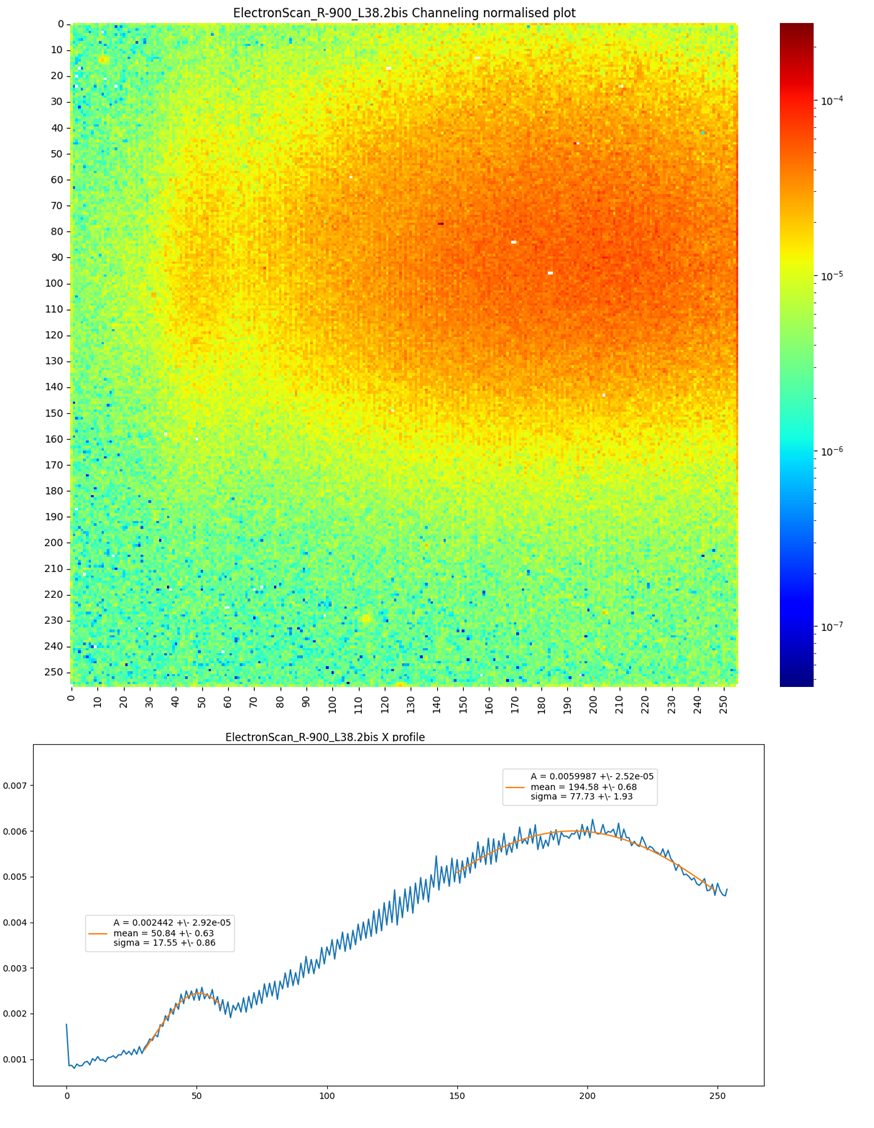}
}
\caption{Normalized 2d Beam spot distribution of an electron Channeling run and corresponding X-projection profile.}
\label{fig:eleCH}       
\end{figure}

\begin{figure}
\centering
\resizebox{0.5\textwidth}{!}{%
  \includegraphics{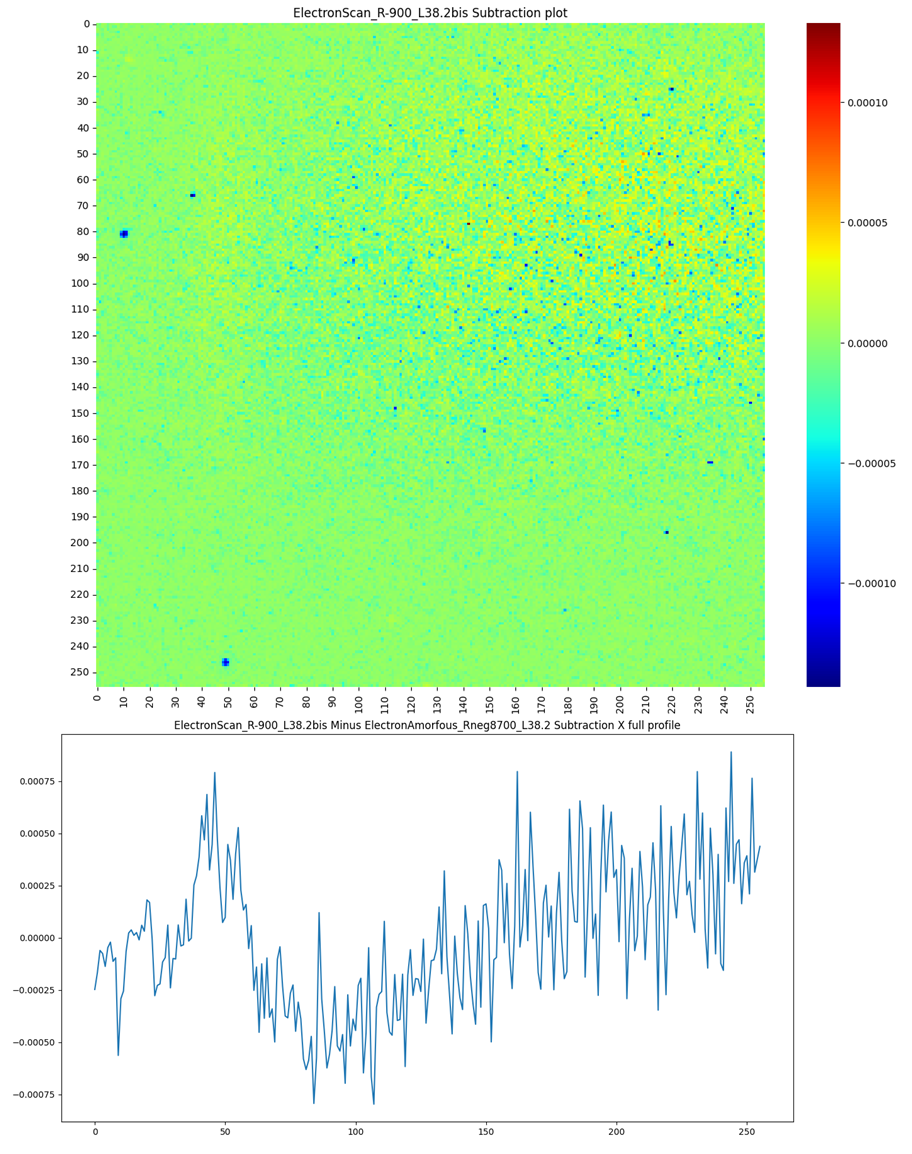}
}
\caption{Subtraction between the normalized electron CH and the AM runs distributions and corresponding X-projection profile. }
\label{fig:eleCH-eleAm}       
\end{figure}

The same kind of measurements and analysis has been performed also with a 450 MeV electron beam, provided just setting the LINAC and the BTF line in electron configuration. As before, the CH peak appears in the 2D particle distribution on the TimePix3 (Fig.\ref{fig:eleCH}). In Fig.\ref{fig:eleCH-eleAm} instead is shown the subtracted profile. As expected, the CH peak is much less intense with respect to the positron case. In fact, negative particles have a lower probability to be canalized in the crystal planes \cite{mazzolari2014steering,wistisen2016channeling}.

Unfortunately, the measurement with electrons has been performed during the last night of measurements and the statistics collected is not enough to show a precise profile and, consequently, to establish the channeling angle, which anyway appears compatible with the positron results.

\section{Conclusions}
\label{sec:Conlcusions}

The experimental results in this paper demonstrate the possibility to deflect 450 MeV positrons beams using a bent silicon crystal for slow extraction applications. 

The most important technological achievements are the production of a crystal thin enough to minimize the DCH effect, and a bending holder able to bend it with a very uniform curvature, obtaining a deflection angle beyond 1 mrad.

This technology has been optimized reaching a record deflection angle of 1.26 mrad, $\sim$ 0.3 mrad more with respect previous results obtained with silicon crystals \cite{mazzolari2014steering,wistisen2016channeling,sytov2017steering,bandiera2021investigation}, without breaking any samples. Two additional silicon crystals are ready for future characterization, with a higher deflection angle expected. Furthermore, this upgrade enables the serial production of these bent crystals. 

An additional important result, however, is the commissioning and the management of the experimental set up for the crystal characterisation, in terms of beam spot and divergence, alignment, but also the development of new measurement and analysis techniques. The possibility to use also not optimized positron beams, with a divergence ten times higher than the PCH acceptance angle, for bent crystal characterization is demonstrated. In fact, secondary sub-GeV positron beams, including the BTF one, are generally not intrinsically suitable for this type of applications. 

Recently, another experiment used the newly developed 530 MeV low divergence beam of Mainz Microtron MAMI, to complete a very detailed study about coherent process of sub-GeV positrons in bent silicon crystals, including PCH \cite{mazzolari2024observation}, strongly supporting the results reported here.

The beam steering of sub-GeV positrons and electrons can find several applications in leptons accelerator physics. The results reported here, scientifically and technologically interesting in themselves, complete also the SHERPA feasibility study of slowly extracting a spill of $O$(ms) from one of the DA$\Phi$NE accelerator rings \cite{garattini2021crystal}.

Moreover, this experimental results will be extremely useful as benchmark for simulation tools.

The next step will be performing additional measurements at the BTF, to characterise the other crystal samples, and, in particular, to complete the electron CH measurements.

\section{ACKNOWLEDGMENTS}
\label{sec:Ackno}

This work has been financially supported by the Istituto Nazionale di Fisica Nucleare (INFN), Italy, Commissione Scientifica Nazionale 5, Ricerca Tecnologica—Bando No. 21188/2019.
The authors would like to express their gratitude to the Frascati National Laboratory and the whole SPCM, BTF and the PADME teams for the strong support provided for these activities.
Finally, the authors wish to acknowledge INFN-Fe and LNL-INFN, for their work on constructing the silicon crystals and to share the bending holder technology.

%
\bibstyle{abbrvant}
\bibliography{all_bib}
%
%
%

\end{document}